\documentclass[12pt]{article}

\usepackage[totalheight = 23cm, totalwidth = 17cm]{geometry}
\usepackage{amssymb,amsmath,amsfonts,amsbsy,graphicx,bm}

\usepackage{color,cancel,ulem, hyperref}

\newcommand{\dis}[1]{\begin{equation}\begin{split}#1\end{split}\end{equation}}

\begin{document}

\begin{titlepage}

\begin{center}

{\LARGE \bf 
Background instability  of quintessence model in light of entropy and distance conjecture
}

\vskip 1.0cm

{\large
Min-Seok Seo$^{a}$ 
}

\vskip 0.5cm

{\it
$^{a}$Department of Physics Education, Korea National University of Education,
\\ 
Cheongju 28173, Republic of Korea
}

\vskip 1.2cm

\end{center}

\begin{abstract}
We apply the covariant entropy bound argument supporting the de Sitter swampland conjecture to the quintessence model, to find out the condition for the background to be unstable.
More concretely, the background is unstable when the matter entropy given by the species number of the effective field theory increases more rapidly than the geometrical entropy proportional to the apparent horizon area, since it contradicts the covariant entropy bound.
The rapid increase in the matter entropy is proposed by the distance conjecture, which states that the time evolution of some scalar field along the geodesic in the field space brings about the descent of a tower of states from UV.
  From this, we find that  for the quintessence model, the unstable background admits the  event horizon of  finite size,   and the converse is also true when the string tower is taken into account.
Here, the presence of the event horizon implies that the trans-Planckian modes can be classicalized, violating the trans-Planckian censorship bound.  
We also point out that the scale separation between the Kaluza-Klein mass scale and the Hubble parameter can be realized when the product between the increasing rates of the matter and   the geometrical entropies is bounded from below, which is consistent with the AdS distance conjecture.
Our study suggests that various swampland conjectures can be comprehensively understood in the language of the entropy.

\end{abstract}

\end{titlepage}

\newpage

\section{Introduction}

In the framework of low energy effective field theory (EFT), cosmological models are often constructed by observing  behavior of scalar fields in the potential.
For example, primordial inflation can be described by the slow-roll of the  inflaton on the nearly flat potential, while the cosmological constant is realized by the stabilization of the moduli at the local minimum of the potential with positive vacuum energy.
At first glance, it appears that at sufficiently low energy scale, such EFT description decouples from its UV completion, namely, quantum gravity.
However, recent swampland program claims that this may not be the case, by conjecturing quantum gravity constraints on the EFT motivated by  observations in string theory \cite{Vafa:2005ui} (for reviews, see, \cite{Brennan:2017rbf, Palti:2019pca, vanBeest:2021lhn, Grana:2021zvf, Agmon:2022thq}).

Intriguingly, some swampland conjectures predict that the background  describing the accelerating expansion of the universe cannot be stable.
According to the de Sitter (dS) swampland conjecture \cite{Danielsson:2018ztv, Obied:2018sgi} (see  also  \cite{Andriot:2018wzk, Garg:2018reu, Cicoli:2018kdo, Ooguri:2018wrx, Andriot:2018mav} for a refinement), the scalar potential consistent with quantum gravity must satisfy either 
\dis{ \frac{|\nabla V|}{\kappa_d V}\sim {\cal O}(1),\quad\quad {\rm or} \quad\quad {\rm min}\Big[\frac{ \nabla \nabla V}{\kappa_d^2 V}\Big]\lesssim -{\cal O}(1),}
where $\nabla$ is taken with respect to scalar fields and $\kappa_d=\sqrt{8\pi G_d}=M_{{\rm Pl},d}^{-\frac{d-2}{2}}$, hence dS space is unstable.
This can be supported by the  covariant entropy bound \cite{Bousso:1999xy, Bousso:2002ju} as well as the distance conjecture \cite{Ooguri:2006in}.
More concretely, the distance conjecture states that in the infinite distance limit of moduli space,  a tower of states, which turns out to be either the Kaluza-Klein (KK) or the string  tower \cite{Lee:2019xtm, Lee:2019wij} becomes light, invalidating the EFT.
This implies that the time evolution of the scalar field like the inflaton  along the geodesic  in the field space  brings about the descent of an associated tower of states from UV.
Then, the entropy inside the apparent horizon increases in time and eventually exceeds the covariant entropy bound proportional to the area of the apparent horizon \cite{Ooguri:2018wrx}.
To avoid this inconsistency, the background   must be deformed such that the apparent horizon area increases in time.
\footnote{  The idea that comparing the matter entropy in the bulk with the geometrical entropy obeying the area law has been discussed earlier in, for example, \cite{Padmanabhan:2012ik, Hadi:2019qxn}.}

On the other hand, the trans-Planckian censorship conjecture proposes the upper bound on the '`lifetime" of the accelerating phase by requiring that quantum fluctuations above the Planck scale cannot be classicalized \cite{Bedroya:2019snp}.
Since the accelerating phase cannot last long enough for the   trans-Planckian  modes to be   stretched outside the horizon, the values of the scale factor $a(t)$ at  initial and final times of the accelerating phase, denoted by $t_i$ and $t_f$, respectively, satisfy
\dis{\frac{a(t_f)}{a(t_i)}\lesssim \frac{M_{{\rm Pl}, d}}{H(t_f)},\label{eq:TCC}}
where $H\equiv\dot{a}/a$ (dot here denotes the derivative with respect to $t$) is the Hubble parameter.
For quasi-dS background where $a(t)\simeq e^{Ht}$ with almost constant $H$, the lifetime $t_f-t_i$ is given by $H^{-1}\log(M_{{\rm Pl},d}/H)$.
In the case of the quintessence model \cite{Peebles:1987ek, Ratra:1987rm, Caldwell:1997ii} (for reviews, see, e.g.,  \cite{Martin:2008qp, Tsujikawa:2013fta})
\footnote{For discussions in the context of string model, see   \cite{Hellerman:2001yi, Fischler:2001yj, Kaloper:2008qs, Cicoli:2012tz} and also \cite{Olguin-Trejo:2018zun, Cicoli:2021fsd, Cicoli:2021skd, Brinkmann:2022oxy, Conlon:2022pnx, Rudelius:2022gbz, Calderon-Infante:2022nxb, Apers:2022cyl, Shiu:2023nph, Shiu:2023fhb, Seo:2024fki, Seo:2024qzf}, which are motivated by the swampland program. }
 where the potential  decays exponentially and the scale factor is given by $a(t)\sim t^p$, hence, $H=p/t$, the bound from the trans-Planckian censorship conjecture reads $p\leq 1$, since $t_i$ is required to be larger than the Planck time $M_{{\rm Pl}, d}^{-1}$.
This is equivalent to the condition that the expansion of the  universe is slow enough not to allow  the event horizon  \cite{Hellerman:2001yi, Fischler:2001yj}, from which  \cite{Bedroya:2022tbh} (see also \cite{Bedroya:2025fie}) claimed that the trans-Planckian censorship conjecture can be supported by the holographic principle.
More concretely,   in the presence of the (lightlike) event horizon, the future infinity is given by the spacelike hypersurface, which may be interpreted as the holographic boundary according to the dS/CFT correspondence \cite{Strominger:2001pn}.
However, the event horizon erases   quantum information through the classicalization.
In particular, the massive modes, including  tower  states, easily lose   nontrivial information when they are stretched beyond the event horizon.
 \footnote{For details, we refer the reader to \cite{Bedroya:2022tbh}, which also provides the following heuristic argument.
 Given the momentum   $|{\mathbf p}|=k/a$  of the massive mode, the comoving velocity in the non-relativistic limit is  $v=k/(m a^2)$.
 Then the maximum distance that the mode can travel given by $\int^\infty v dt$  is convergent (so cannot reach the future infinity) when   $|{\mathbf p}|>1/2$, that is, even if the value of $p$ is not as large as $1$, hence, the universe does not expand fast enough to have the event horizon.
 }
 Therefore,  the future infinity encodes information on the frozen correlators only hence cannot describe  the bulk degrees of freedom   completely, which contradicts the holographic principle.

 It is remarkable that the holographic argument for the trans-Planckian censorship conjecture is similar  to the covariant entropy bound argument for the dS swampland conjecture in the situation considered : in both cases, the problem arises when the entropy we can read off from the geometry cannot count some of the matter degrees of freedom, in particular those from a rapidly increasing number of tower states.
 This motivates us to investigate the connection between two conjectures more carefully, in particular from the point of view of entropy as well as the distance conjecture. 
 One immediate way to approach this issue is to compare the lifetime of the accelerating phase obtained by the covariant entropy bound argument for  the dS swampland conjecture  with that predicted by the trans-Planckian censorship conjecture.
 For quasi-dS space, the former is estimated as $(\sqrt{\langle\epsilon_H \rangle} H)^{-1}\log(M_{{\rm Pl}, d}/H)$ where $\langle\epsilon_H \rangle$ is the mean value of the slow-roll parameter $\epsilon_H\equiv -\dot{H}/H^2$ over the corresponding time scale  \cite{Seo:2019wsh, Cai:2019dzj} (see also \cite{Sun:2019obt, Cribiori:2025oek}).
 This ``entropic quasi-dS instability time" is more or less consistent with the lifetime predicted by the trans-Planckian censorship conjecture, except for the enhancement by $1/\sqrt{\langle\epsilon_H \rangle}$.
 Indeed, such enhancement appears naturally because the distance travelled by the scalar in the field space during $\Delta t$ is $\dot{\phi}\Delta t$ and $\dot{\phi}$ is proportional to $\sqrt{\epsilon_H}$ by the equations of motion (see \eqref{eq:EoM}).
 \footnote{Intuitively, the nonzero $\epsilon_H$ (hence $\dot{H}$) indicates that the timelike dS  isometry is spontaneously broken, so $\dot{\phi}$ does not vanish.
 This means that different time slices are physically distinct, which can be labelled by the value of $\phi(t)$ as a function of time. 
   }
 While $\epsilon_H$ is smaller than $1$, if the potential is sufficiently flat only over the small field range, $1/\sqrt{\langle\epsilon_H \rangle}$ may be close to ${\cal O}(1)$.

 On the other hand,  the exponentially decaying potential in the quintessence model is ubiquitous in string theory model  where the large moduli limit is taken for the perturbative control \cite{Hellerman:2001yi, Fischler:2001yj, Kaloper:2008qs, Cicoli:2012tz}.
 Then, one may try to apply the covariant entropy bound argument to the quintessence model to obtain the instability condition of the background as well as the lifetime of the accelerating phase and compare them with  the trans-Planckian censorship conjecture, which is the main topic of this article.
 For self-containedness, we begin discussion in section \ref{Sec:Rev} with the review as follows.
 Section \ref{subsec:distance} summarizes how the distance conjecture is realized  in string theory when the associated cutoff scale is given by the species scale,  and section \ref{subsec:cosmo} collects the solutions to the equations of motion for the scale factor and the scalar field in the quintessence model.
 Based on this, in section  \ref{Sec:Entropy}, we compare the geometrical entropy we can read off from the apparent horizon area and the matter entropy contributed by the descending   tower states.
  As for the matter entropy, \cite{Ooguri:2018wrx} considered an ansatz motivated by the entropy of particles in the thermal bath, which scales more rapidly than the area in time.
  Meanwhile,  it has been suggested that the species number under the cutoff scale can be regarded as the entropy of the EFT itself  \cite{Cribiori:2023ffn, Basile:2023blg, Basile:2024dqq, Herraez:2024kux, Herraez:2025clp}.
  Explicit expressions of this ``species entropy" as well as  the geometrical entropy are given in section \ref{Sec:entropyreview}.
 As discussed in detail in  section \ref{Sec:instab}, since the  species entropy can be interpreted as  the entropy of the smallest black hole in the EFT, hence relevant to the   UV scale, it is required to be smaller than the geometrical entropy (which is relevant to the IR scale) for the background to be stable.
 From this,  one finds that if the background is unstable, that is, the species entropy  increases rapidly and eventually   exceeds the geometrical entropy, the background geometry has the event horizon, violating the trans-Planckian censorship conjecture. 
Then, we   obtain the lifetime of such unstable  background.
Moreover, since the geometrical entropy and the species entropy we consider are closely connected to  $H$ and the tower mass scale, respectively, we can discuss the implication of our results on the scale separation \cite{Coudarchet:2023mfs}, which will be addressed  in section \ref{Sec:scale}.
 After summarizing discussions, we conclude.

 \section{Distance conjecture  and quintessence model : review}
 \label{Sec:Rev}
 
 \subsection{Distance conjecture in string theory and   species scale}
 \label{subsec:distance}
 
To begin with, we consider the quantum gravity cutoff  of the $d$-dimensional EFT  which is appropriate to describe the distance conjecture.
 Since the graviton universally couples to all the particle species, the  gravitational interaction in the loop corrections becomes strong  above the ``species scale"
 \dis{\Lambda_{\rm sp}=\frac{M_{{\rm Pl}, d}}{N_{\rm sp}^{\frac{1}{d-2}}},\label{eq:OrDef} }
 where $N_{\rm sp}$ is the number of particle species below $\Lambda_{\rm sp}$.
 From this, it has been claimed that the natural quantum gravity cutoff is not $M_{{\rm Pl},d}$, but the lower scale   $\Lambda_{\rm sp}$   \cite{Veneziano:2001ah, Dvali:2007hz,Dvali:2007wp, Dvali:2009ks, Dvali:2010vm, Dvali:2012uq}.
 In particular, when $N_{\rm sp}$ is dominated by $n$ towers of states with tower mass scale $m_t$, it can be approximated as $N_{\rm sp}=(\Lambda_{\rm sp}/m_t)^n$, which gives
 \dis{\Lambda_{\rm sp}=\Big(\frac{m_t}{M_{{\rm Pl}, d}}\Big)^{\frac{n}{d+n-2}}M_{{\rm Pl}, d},\quad\quad
 N_{\rm sp}=\Big(\frac{M_{{\rm Pl}, d}}{m_t}\Big)^{\frac{n(d-2)}{d+n-2}}.\label{eq:specLN}}
 These expressions are consistent with the distance conjecture : the decrease in $m_t$ leads to the increase in $N_{\rm sp}$, hence, the decrease in $\Lambda_{\rm sp}$.
 
 In the string theory model, a tower of states is either the KK or the string tower, which realizes the distance conjecture as follows.
 First, the properties of the KK tower are determined by the stabilized values  of the K\"ahler moduli   such as the volume modulus $u(x)$. 
 To see this quantitatively, we consider the $(d+n)$-dimensional spacetime  decomposed into the $d$-dimensional noncompact spacetime (with coordinates $x^\mu$) and the  $n$-dimensional compact space (with coordinates $y^m$), with the metric ansatz 
\dis{ds^2&=e^{-\frac{2n}{d-2} u(x)}g_{\mu\nu}dx^\mu dx^\nu +e^{2u(x)}g_{mn}dy^m dy^n
\\
&=e^{-\frac{2n}{d-2} u(x)}\big(g_{\mu\nu}dx^\mu dx^\nu +e^{\frac{2(d+n-2)}{d-2}u(x)}g_{mn}dy^m dy^n\big).}
The first expression shows  that $u(x)$ is interpreted as the volume modulus in the sense that the volume of the compact space is given by $e^{n u}\int d^n y\sqrt{g_n}$, and the second expression shows that the KK mass scale is given by exp$[-\frac{d+n-2}{d-2}u ] M_{{\rm Pl}, d}$.
Then, under the dimensional reduction, the $(d+n)$-dimensional Einstein-Hilbert action becomes the $d$-dimensional effective action,
{\small
\dis{S_d=\frac{1}{2\kappa_d^2}\int d^d x \sqrt{-g_d}\Big[{\cal R}_d+e^{-2\frac{d+n-2}{d-2}u}{\cal R}_n+\frac{2n}{d-2}g^{\mu\nu}\nabla_\mu\nabla_\nu u-\frac{n(d+n-2)}{d-2}g^{\mu\nu}\nabla_\mu u \nabla_\nu u\Big],\label{eq:Sd}}
}
where  ${\cal R}_d$ and ${\cal R}_n$ are Ricci scalars for the $d$-dimensional noncompact spacetime and $n$-compact space, respectively, and 
\dis{\frac{1}{\kappa_d^2}=\frac{1}{8\pi G_d}=M_{{\rm Pl},d}^{d-2}=\frac{1}{\kappa_{d+n}^2}\int d^ny\sqrt{g_n}.}
From this, one finds that the canonically normalized volume modulus $\rho(x)$ is given by
\dis{\rho(x)=\frac{1}{\kappa_d}\sqrt{\frac{n(d+n-2)}{d-2}}u(x),}
in terms of which the KK mass scale and the potential generated by ${\cal R}_n$ (the second term in \eqref{eq:Sd}) are written as
\dis{&m_{\rm KK}=e^{-\frac{d+n-2}{d-2}u } M_{{\rm Pl}, d} = e^{- \sqrt{\frac{d+n-2}{n(d-2)}}\kappa_d\rho } M_{{\rm Pl}, d},
\\
&V_{\cal R}=e^{-2\frac{d+n-2}{d-2}u }{\cal R}_n = e^{-2 \sqrt{\frac{d+n-2}{n(d-2)}}\kappa_d \rho }{\cal R}_n,\label{eq:mkkVr}}
respectively.
We also infer from \eqref{eq:specLN} that $\Lambda_{\rm sp}$ and $N_{\rm sp}$ associated with the KK tower (that is, $m_t=m_{\rm KK}$) can be written as 
\dis{\Lambda_{\rm sp}=M_{{\rm Pl}, d} e^{-\sqrt{\frac{n}{(d-2)(d+n-2)}}\kappa_d \rho },
\quad\quad
N_{\rm sp}=e^{\sqrt{\frac{n(d-2)}{d+n-2}}\kappa_d \rho }.\label{eq:KKLN}}
Therefore, as stated by the distance conjecture, $m_{\rm KK}$   decreases exponentially in the increasing direction of $\rho$, resulting in the rapid increase in $N_{\rm sp}$, hence, the rapid decrease in the cutoff  $\Lambda_{\rm sp}$,  which is  nothing more than the $(d+n)$-dimensional Planck scale.

Meanwhile, the scalar field associated with   the string tower is the dilaton $\phi(x)$.
To see this, we consider the $d$-dimensional action in the string frame,
\dis{S_{d,s}=\frac{1}{2\tilde{\kappa}_d^2}\int d^d x \sqrt{-g_{d, (s)}}e^{-2\phi}\big[{\cal R}_d+4g_{(s)}^{\mu\nu}\nabla_\mu \phi \nabla_\nu \phi\big]+\cdots,}
where $\tilde{\kappa}_d^{2}=(4\pi M_s^{d-2})^{-1}$.
\footnote{If the $d$-dimensional spacetime is obtained by compactifying the $(d+n)$-dimensional one on the $n$-dimensional compact space, the volume of which is given by ${\cal V}_nM_s^{-n}$, since $\tilde{\kappa}_{d+n}^{-2}=4\pi M_s^{d+n-2}$, we obtain $\tilde{\kappa}_d^{-2}=4\pi M_s^{d-2} {\cal V}_n$.
Then, ${\cal V}_n$ can be absorbed in the definition of $d$-dimensional dilaton $\phi$ through the relation $e^{-2\phi}=e^{-2\Phi}{\cal V}_n$ where $\Phi$ is the  $(d+n)$-dimensional dilaton.}
Performing the Weyl rescaling ${g_{(s)}}_{\mu\nu}=e^{\frac{4}{d-2}(\phi-\phi_0)}g_{\mu\nu}$, we get the Einstein frame action 
\dis{S_{d,E}=\frac{1}{2{\kappa}_d^2}\int d^d x \sqrt{-g_d}\Big[{\cal R}_d-4\frac{d-1}{d-2}g^{\mu\nu}\nabla_\mu\nabla_\nu\phi-\frac{4}{d-2}g^{\mu\nu}\nabla_\mu \phi\nabla_\nu \phi\Big],}
where 
\dis{\kappa_d^{2}=\tilde{\kappa}_d^2 e^{2\phi_0}=\frac{e^{2\phi_0}}{4\pi M_s^{d-2}},}
and $e^{\phi_0}$ is identified with the string coupling constant $g_s$.
Then, one finds that the canonically normalized dilaton field is given by
\dis{D(x)=\frac{1}{\kappa_d}\frac{2}{\sqrt{d-2}}\phi(x),}
and given fixed value of $\kappa_d$ (hence $M_{{\rm Pl},d}$), the string mass scale satisfies
\dis{M_s=\frac{M_{{\rm Pl},d}}{(4\pi)^{\frac{1}{d-2}}}e^{\kappa_d\frac{1}{\sqrt{d-2}}D},}
where $D$  indicates the stabilized value of the dilaton, that is, the value of $D$ when $\phi=\phi_0$.
Now we note that whereas the mass of the string excitation is given by $m_n=\sqrt{n}M_s$ ($n \in \mathbb{Z}^{0+}$), the number of degeneracy grows exponentially as $d_n\sim e^{\sqrt{n}}$ \cite{Kani:1989im}, indicating that $\Lambda_{\rm sp}$ and $N_{\rm sp}$ for the string tower correspond to the $n\to \infty$ limit of relations in \eqref{eq:specLN} \cite{Castellano:2021mmx} 
\dis{\Lambda_{\rm sp}=M_s =\frac{M_{{\rm Pl},d}}{(4\pi)^{\frac{1}{d-2}}}e^{\kappa_d\frac{1}{\sqrt{d-2}}D},\quad\quad
N_{\rm sp}=\Big(\frac{M_{{\rm Pl}, d}}{M_s}\Big)^{d-2}=4\pi e^{-\kappa_d \sqrt{d-2}D}.\label{eq:stLN}}
These are also consistent with the distance conjecture : in the limit $D\to -\infty$, where $g_s$ becomes extremely tiny, a tower of string excitation descends from UV such that $\Lambda_{\rm sp}$, which is in fact $M_s$  decreases exponentially.

 As we have seen, various quantities like $m_t$, $\Lambda_{\rm sp}$, and $N_{\rm sp}$ in the string theory model depend on some scalar field value, namely,  $\rho$ for the KK tower and $D$ for the string tower,  realizing  the distance conjecture.
 Denoting such scalar dependent quantity by $F$ comprehensively, the rate of change with respect to the scalar field value is written as $\nabla F/F$ where $\nabla$ is taken with respect to the scalar field.
 As pointed out in \cite{Castellano:2023stg, Castellano:2023jjt}, for both the KK and the string tower, the product of the decreasing rate of   $m_t$ ($m_{\rm KK}$ for the KK tower and $M_s$ for the string tower) and that of $\Lambda_{\rm sp}$  is independent of   details of the tower, for example, the value of $n$, but determined by $d$ only  
 \dis{\frac{\nabla m_t}{m_t}\cdot \frac{\nabla \Lambda_{\rm sp}}{\Lambda_{\rm sp}} =\frac{1}{d-2}.}

\subsection{Quintessence model}
\label{subsec:cosmo}

On the other hand, in the framework of the low energy EFT, the $d$-dimensional homogeneous and isotropic universe can be realized by the scalar-gravity system, where  the background geometry is described by the  Friedmann-Lema\^itre-Robertson-Walker (FLRW) metric  $ds^2=-dt^2+a(t)^2 dx^idx^i$ and $\phi(t)$ depends only on $t$.
 They correspond to the solutions of the following equations of motion  :
 \dis{&\frac{(d-1)(d-2)}{2 \kappa_d^2}H^2=\frac12\dot{\phi}^2+V(\phi),
 \\
 &\dot{H}=-\frac{\kappa_d^2}{d-2}\dot{\phi}^2,
 \\
 &\ddot{\phi}+(d-1)H\dot{\phi}+\frac{dV}{d\phi}=0.\label{eq:EoM}}
 In particular, in  the quintessence model, the potential decays exponentially as 
 \dis{V=V_0 e^{-\lambda \kappa_d\phi},}
  where $\lambda=-\nabla_\phi V/(\kappa_d V)$ is the decreasing rate of the potential with respect to the modulus $\phi$. 
In this case,  one finds two types of   attractor solutions, which correspond to the  classical configurations of $a(t)$ and $\phi(t)$ in the limit $t\to\infty$ \cite{Rudelius:2022gbz, Bedroya:2022tbh} (see also \cite{Lucchin:1984yf, Copeland:1997et, vandenHoogen:1999qq} for earlier discussion),
 \dis{{\bf Solution~A}  :\quad \kappa_d \phi(t)=\sqrt{\frac{d-2}{d-1}}\log \Big(\frac{t}{t_0}\Big),\quad\quad
a(t)=a_0\Big(\frac{t}{t_0}\Big)^{\frac{1}{ d-1 }}.\label{eq:solA}}
and
\dis{{\bf Solution~B}  :\quad \kappa_d\phi(t)=\frac{2}{\lambda}\log\Big[\sqrt{\frac{\kappa_d^2\lambda^2 V_0 }{2\big(\frac{4}{\lambda^2}\frac{d-1}{d-2}-1\big)}}t \Big],\quad\quad
a(t)=a_0\Big(\frac{t}{t_0}\Big)^{\frac{4}{(d-2)\lambda^2}}.\label{eq:solB}} 
We note that the solution B exists only if $\lambda<2\sqrt{\frac{d-1}{d-2}}$.
  For this condition to be conceivable, we restrict our discussion to the case of $d>2$ only. 

In both cases, the scale factor depends on time as $a(t)\sim t^p$ for some real number $p$, then the radius of the event horizon,
 \dis{\int_t^\infty \frac{dt'}{a(t')}=\frac{1}{1-p}{t'}
 ^{1-p}\Big|_t^\infty}
 is finite provided $p>1$.
 \footnote{When $p=1$, the event horizon radius diverges logarithmically.}
 This is consistent with the intuition that for large value of $p$ the universe is close to the accelerating phase, in  which a static observer is causally connected  with a finite region only, implying the finite size of   an event horizon.
 Moreover, the discussion below \eqref{eq:TCC} shows that the condition of the finite size of the event horizon is equivalent to the condition that the trans-Planckian censorship conjecture is violated.
 For the solution A, the exponent $p=\frac{1}{d-1}$ is always smaller than $1$ provided $d>2$, which we will assume throughout the discussion.
 Then, the size of the event horizon is infinite; hence,   the whole region of the universe is causally connected.
 Indeed, as can be inferred from the fact that the solution does not depend on $\lambda$, the value of $V(\phi)$ in this case is negligibly small such that the density-to-pressure ratio is close to $1$.
 This cannot describe the accelerating universe.
 On the other hand,   solution B allows the accelerating phase : when $\lambda<\frac{2}{\sqrt{d-2}}$, $p=\frac{4}{(d-2)\lambda^2}$ can be larger than $1$, indicating the finite    size of the event horizon.
  This shows that  the accelerating universe is realized for small value of $\lambda$, which is evident from the fact that the potential in 
  the limit $\lambda\to 0$ corresponds to the cosmological constant realizing dS space.
  It is also notable that when the potential is given by $V_{\cal R}$ in \eqref{eq:mkkVr}, the value of $\lambda$ which will be denoted by $\lambda_{\cal R}$,
  \dis{\lambda_{\cal R}=2 \sqrt{\frac{d+n-2}{n(d-2)}},\label{eq:lR}}
  is larger than $\frac{2}{\sqrt{d-2}}$ (corresponding to $n\to \infty$) and smaller than $2\sqrt{\frac{d-1}{d-2}}$ (corresponding to $n =1$), the upper bound on $\lambda$ for the solution B.
Moreover, if the potential is dominated by the potential with $\lambda > \lambda_{\cal R}$, the event horizon cannot be finite.

 \section{ Entropy argument for the quintessence model}
 \label{Sec:Entropy}
 
\subsection{Matter entropy and geometrical entropy}
\label{Sec:entropyreview}

 The dS swampland conjecture claiming the instability of (quasi-)dS space can be supported by the covariant entropy bound argument \cite{Ooguri:2018wrx}.
 That is, a descent of a tower of states as predicted by the distance conjecture leads to the rapid increase in   the matter entropy inside the apparent horizon, which eventually exceeds the geometrical entropy proportional to the apparent horizon area.
 This contradicts   the covariant entropy bound,   unless the  spacetime background   is deformed such that the apparent horizon area also increases rapidly enough to accommodate the increasing matter entropy.
 This article is devoted to apply this argument to the quintessence model.
 
The form of the matter entropy in \cite{Ooguri:2018wrx} was assumed  on dimensional ground   (for an attempt to interpret this ansatz, see, e.g., \cite{Seo:2019mfk}).
This is motivated by the form of the entropy of particles in the thermal bath : it scales as the volume, hence even if small initially, it eventually exceeds  the geometrical entropy scaling as the area  when the universe expands rapidly.
  The implication of the distance conjecture on the matter entropy scaling as the volume was discussed in, for example,  \cite{Seo:2022uaz}.

  On the other hand, from the observation that $N_{\rm sp}$ increases exponentially with respect to the modulus value (see, e.g., \eqref{eq:KKLN} and \eqref{eq:stLN}), it has been argued that the states in the moduli space can be described like the thermodynamic system, and in this  ``species thermodynamics", $N_{\rm sp}$ is interpreted as the entropy of the EFT moduli space   \cite{Cribiori:2023ffn, Basile:2023blg, Basile:2024dqq, Herraez:2024kux, Herraez:2025clp} (see also \cite{vandeHeisteeg:2022btw, Cribiori:2022nke}).
   An intuitive  (but not so precise)  way to see this is to observe that when most of tower states are well below   the species scale, we may treat $N_{\rm sp}$ particle species as  massless ones in the leading approximation.
    Then,   there can be $N_{\rm sp}!$ states made up of  $N_{\rm sp}$ particle species, from which the entropy is estimated as    $S_{\rm sp}\simeq \log(N_{\rm sp}!)\simeq N_{\rm sp}(\log N_{\rm sp}-1)$, which is not much enhanced from $N_{\rm sp}$ when compared with, for example, another polynomial like $N_{\rm sp}^2$.
    More rigorous treatment in \cite{Basile:2023blg}  shows that $S_{\rm sp}$ is indeed given by $N_{\rm sp}$ when $N_{\rm sp}$ is very large. 
 More rigorously , given a tower of states with mass  $m_n = f(n) m_t$ and degeneracy $d_n$, the tower state in the highest level can be labelled by an integer $N$ satisfying $\Lambda_{\rm sp}=f(N) m_t$ and $\sum_{n \leq N}d_n=N_{\rm sp}$.
 Then, the energy, or equivalently, the mass of the smallest black hole made up of all the particle species is given by $E=\sum_{n \leq N}d_n m_n=\sum_{n \leq N}d_n f(n) m_t$, which will be denoted by $E=M m_t$, where $M=\sum_{n \leq N}d_n f(n)$.
 Now, we define the auxiliary partition function,
  \dis{Z(q)=\sum_M D(M) q^M=\prod_{n \leq N}\frac{1}{(1-q^{f(n)})^{d_n}},}
  from which one obtains ``species entropy" $S_{\rm sp}=\log D(M)$ with
   \dis{D(M)=\frac{1}{2\pi i}\oint\frac{dq}{q^{M+1}}Z(q).} 
 The integral can be computed by the steepest descent method. 
 Denoting the value of $q$ corresponding to the saddle point of $-(M+1)\log q +\log Z(q)$ by $q_*$, since $q_*$ must be close to $1$,  we can set $q_*=1-\varepsilon$, where $\varepsilon$ is obtained from
  \dis{0&=\frac{d}{dq}\Big[-(M+1)\log q +\log Z(q)\Big]_{q=q_*}=-\frac{M+1}{q_*}+\frac{1}{q_*}\sum_{n\leq N} d_n f(n)\frac{q_*^{f(n)}}{1-q_*^{f(n)}}
  \\
  &\simeq  -\frac{M}{q_*}+\frac{1}{q_*}\sum_{n\leq N} d_n f(n)\frac{1}{f(n)\varepsilon}=-\frac{M}{q_*}+\frac{1}{q_*\varepsilon}N_{\rm sp}.}
Here, for the last equality, $\sum_{n \leq N}d_n=N_{\rm sp}$ is used.
  Therefore,  $q_*\simeq 1-(N_{\rm sp}/M)$, and putting this back into the integral, one finds that the leading term of $S_{\rm sp}=\log D(M)$ is given by $N_{\rm sp}$.
  Then, the contribution of a tower of states to $N_{\rm sp}$ given by \eqref{eq:specLN} can be interpreted as that to $S_{\rm sp}$.
  
  For $S_{\rm sp}$ to be interpreted as the entropy in the real world, in particular, in cosmology, there needs to be a macroscopic system, which can be realized in $D(M)$ microscopic ways.
  In species thermodynamics, the macroscopic system is given by the black hole of the radius $\Lambda_{\rm sp}^{-1}$ (hence temperature $\Lambda_{\rm sp}$).
  This can be supported by the fact that $S_{\rm sp}$ obeys the area law as $S_{\rm sp}=N_{\rm sp}\sim \Lambda_{\rm sp}^{-(d-2)}$.
  Since $\Lambda_{\rm sp}$ is the UV cutoff scale of the EFT,  $\Lambda_{\rm sp}^{-1}$ would be the smallest radius of the black hole we can consider in the framework of the EFT.

  Meanwhile, although the FLRW universe may not have the event horizon, the apparent horizon can be always defined.
  From this,  it has been shown that the Einstein equation is equivalent to the first law of thermodynamics, where the geometrical entropy is given by the  apparent horizon area divided by $4G_d$ \cite{Hayward:1997jp, Bak:1999hd, Cai:2005ra, Cai:2006rs, Abreu:2010ru, Binetruy:2014ela}.
 Here, the apparent horizon is the boundary of the ``trapped surface" (the boundary between the region  where the geodesic congruences contract  and the region where they extend),   the expansion parameter vanishes on the apparent horizon.
 For   the FLRW metric, which can be written as $ds^2=h_{ab}dx^a dx^b+\tilde{r}^2 d\Omega_{d-2}^2$ where $x^a=t, r$ and $\tilde{r}=a(t)r$, this condition becomes $h^{ab}\partial_a \tilde{r}\partial_b \tilde{r}=0$ at $\tilde{r}=\tilde{r}_{\rm app}$, the apparent horizon radius.
  This leads to $\tilde{r}_{\rm app}=a(t)r_{\rm app}=1/H$, hence the geometrical entropy of the apparent horizon is given by
\dis{S_{\rm hor}=\frac14 \frac{8\pi }{\kappa_d^2}\frac{2\pi^{\frac{d-1}{2}}}{\Gamma\big(\frac{d-1}{2}\big)}\frac{1}{H^{d-2}}
= \frac{2\pi }{\kappa_d^2}\frac{2\pi^{\frac{d-1}{2}}}{\Gamma\big(\frac{d-1}{2}\big)}\Big(\frac{t}{p}\Big)^{d-2}.\label{eq:Shor}}  
Here, $t$ and $\phi$ are related by \eqref{eq:solA} or \eqref{eq:solB}, showing that both $\phi$ and $S_{\rm hor}$ increase in time.

  We note that it is typical to interpret $H$ as the IR cutoff scale of the EFT. 
In this regard, the condition $S_{\rm sp}\sim \Lambda_{\rm sp}^{-(d-2)} < S_{\rm hor}\sim H^{-(d-2)}$ can be naturally imposed for the validity of the EFT since the UV cutoff scale $\Lambda_{\rm sp}$ of the EFT must be higher than the IR cutoff scale $H$  (see, e.g., discussion in section 3.2 of \cite{Cribiori:2023ffn}).
Another way to see this condition is to notice that $S_{\rm sp}$ can be interpreted as the entropy of the smallest black hole of the radius $\Lambda_{\rm sp}^{-1}$.
Then, we can consider  the black hole produced by the matter fluctuation, instead of the particles in  the thermal bath. 
For this purpose, we note that the fluctuation of the background is closely connected to that of the density, which is obvious from the Einstein equation, describing the equivalence between them.
 When the fluctuation of the background, hence that of the density is large enough, the  black hole can be produced, which is well defined only if the black hole radius is smaller than $1/H$ : one way to see this is to notice that the 4-dimensional Schwarzschild-dS black hole solution $ds^2=-f(r)dt^2+dr^2/f(r)+r^2d\Omega_2^2$ with $f(r)=1-(2GM/r)-H^2r^2$ is sensible, i.e., has both black hole and cosmological horizons provided $0<2GM<(2/3^{3/2})H^{-1}$. 
 Since both the entropies of the black hole and the cosmological horizon obey the area law,   this condition may be interpreted that   if the black hole radius becomes larger than $1/H$, hence the black hole entropy becomes larger than the geometrical entropy,  the  background cannot accommodate sufficient information that the matter fluctuation can produce.
 This is exactly what we want to avoid by imposing the entropy argument for the dS swampland conjecture.
 Now,  the smallest size of the black hole is given by $\Lambda_{\rm sp}^{-1}$.
 Then the  entropy of this black hole counting the number of possible states forming the black hole    is   $S_{\rm sp}=N_{\rm sp}\sim \Lambda_{\rm sp}^{-(d-2)}$, and the discussion above tells us that this cannot be larger than $S_{\rm hor}$.
 In other words, $S_{\rm sp}$ is the matter entropy for the smallest black hole which must be smaller than $S_{\rm hor}$.
 Therefore, if $S_{\rm sp}\sim N_{\rm sp}$ increases rapidly as predicted by the distance conjecture, the background must be deformed such that $H^{-1}$ becomes larger hence the black hole solution can be defined. 
 \footnote{We also note that there is a conjecture that the concentration of the matter density inside the cosmological horizon must be small enough not to produce the black hole, which is known as the Cohen-Kaplan-Nelson (CKN) bound \cite{Cohen:1998zx}.
 For discussions on the CKN bound in the  accelerating background, we refer the reader to \cite{Banks:2019arz} and also \cite{Seo:2021bpb}, where the implication of the distance conjecture is also discussed.}

\subsection{Instability of the quintessence model in the accelerating phase}
\label{Sec:instab}
  
 We now compare  $S_{\rm sp}$ as the matter entropy contributed by a tower of states with $S_{\rm hor}$ as the geometrical entropy in the quintessence model.
 As time goes on, or equivalently, the value of $\phi$ increases, both $S_{\rm sp}$ and $S_{\rm hor}$ increase.
If the increasing rate of $S_{\rm sp}$ is larger than that of $S_{\rm hor}$, although $S_{\rm sp}< S_{\rm hor}$ initially,  $S_{\rm sp}$ will eventually exceed $S_{\rm hor}$.
This can be interpreted that in the large field limit the geometrical entropy fails to count the increasing number of tower states.
Then, from the same logic as the dS swampland conjecture, we conclude that the spacetime background is unstable.

 We first consider the increasing rate of $S_{\rm sp}=N_{\rm sp}$.
 For the KK mode where $\phi$ is given by  $\rho$,  the increasing rate of the matter entropy can be obtained by using \eqref{eq:KKLN} : 
  \dis{R_{\rm sp}\equiv \Big|\frac{\nabla_\phi S_{\rm sp}}{\kappa_d S_{\rm sp}}\Big|=\Big|\frac{\nabla_\rho N_{\rm sp}}{\kappa_d N_{\rm sp}}\Big| =\sqrt{\frac{n(d-2)}{d+n-2}}=\frac{2}{\lambda_{\cal R}},\label{eq:Rsp}}
  where $\lambda_{\cal R}$ is the decreasing rate of $V_{\cal R}$ given by \eqref{eq:lR}. 
  In the case of the string mode, $\phi=-D$, then from \eqref{eq:stLN}, we obtain $R_{\rm sp}=\sqrt{d-2}$, which corresponds to the $n\to\infty$ limit of \eqref{eq:Rsp}.
Therefore,  we use the expression \eqref{eq:Rsp} as the  increasing rate of $S_{\rm sp}$ for both the KK and the string tower by including the $n\to\infty$ limit.
We also note that
\dis{\Big|\frac{\nabla_\phi \Lambda_{\rm sp}}{\kappa_d\Lambda_{\rm sp}}\Big|=\frac{2}{(d-2)\lambda_{\cal R}}=\frac{R_{\rm sp}}{d-2},}
 which reflects the fact that $S_{\rm sp}=N_{\rm sp}\sim \Lambda_{\rm sp}^{-(d-2)}$, that is, $S_{\rm sp}$ is proportional to the boundary area of the region of  size $1/\Lambda_{\rm sp}$.
  While this relation is a result of \eqref{eq:OrDef}, which reflects the fact that the gravitational coupling is given by $1/M_{{\rm Pl},d}^{d-2}$, it also can be interpreted that the minimal size of the black hole in the framework of the EFT is given by the inverse of the cutoff scale $1/\Lambda_{\rm sp}$, which gives the entropy $S_{\rm sp}$ \cite{Dvali:2007hz}.
  Such an interpretation became the basis of the species thermodynamics \cite{Cribiori:2023ffn}.
  \footnote{This shows the UV nature of $S_{\rm sp}$, which may be relevant to its independence of the IR scale $H$.}

In order to obtain the increasing rate of $S_{\rm hor}$ in the large field limit, we consider two attractor solutions in section \ref{subsec:cosmo}.
Here, we assume that the cosmological evolution is driven by $\rho$ or $D$  such that they can be identified with $\phi$ in \eqref{eq:solA} and \eqref{eq:solB}.
  For the solution A, we infer from \eqref{eq:solA}  that
   \dis{t=t_0e^{\sqrt{\frac{d-1}{d-2}}\kappa_d\phi},}
 or
   \dis{H=\frac{1}{(d-1)t}=\frac{1}{(d-1)t_0} e^{-\sqrt{\frac{d-1}{d-2}}\kappa_d\phi}.\label{eq:HphiA}}
 Comparing this with \eqref{eq:Shor},  we obtain
 \dis{R_{\rm hor}\equiv  \Big|\frac{\nabla_\phi S_{\rm hor}}{\kappa_d S_{\rm hor}}\Big|=\sqrt{(d-1)(d-2)}.}
Then, the ratio
  \dis{\frac{R_{\rm sp}}{R_{\rm hor}}=\sqrt{\frac{n}{(d+n-2)(d-1)}}}
  is smaller than $1$ for any value of $d$ larger than $2$.
  It means that no matter how long we wait, $S_{\rm sp}$ never exceeds $S_{\rm hor}$ provided $S_{\rm sp}<S_{\rm hor}$ initially.

 In contrast, for the solution B, one finds from \eqref{eq:solB} that
 \dis{t=\sqrt{\frac{2\big(\frac{4}{\lambda^2}\frac{d-1}{d-2}-1\big)}{\kappa_d^2 \lambda^2 V_0}}e^{\frac{\lambda}{2}\kappa_d \phi},\label{eq:tphi}}
 or
 \dis{H=\frac{4}{(d-2)\lambda^2 }\frac{1}{t}=\frac{4}{(d-2)} \sqrt{\frac{\kappa_d^2 V_0 }{2\lambda^2\big(\frac{4}{\lambda^2}\frac{d-1}{d-2}-1\big)}} e^{-\frac{\lambda}{2}\kappa_d\phi}, \label{eq:Hphi}}
  showing that the scaling behavior of $H$ with respect to $\phi$ is the same as that of $\sqrt{V}$.
 Since the apparent horizon radius scales as exp$[\frac{\lambda}{2}\kappa_d\phi]$, we obtain
 \dis{R_{\rm hor}=(d-2)\frac{\lambda}{2},}
 (see also  \eqref{eq:Shor}).
 Here, a factor $(d-2)$ comes from the area law of $S_{\rm hor}$.
 Therefore, the ratio is given by
 \dis{\frac{R_{\rm sp}}{R_{\rm hor}}=\frac{4}{(d-2)\lambda \lambda_{\cal R} }=\frac{4}{(d-2)\lambda_{\cal R}^2}\frac{\lambda_{\cal R}}{\lambda},\label{eq:RATIO}}
 indicating that $R_{\rm sp}$ is larger than $R_{\rm hor}$ provided 
 \dis{\lambda< \frac{4}{d-2} \frac{1}{\lambda_{\cal R}}.\label{eq:Cond1}}
 We note that the right-hand side also can be written as
 \dis{ \frac{4}{d-2} \frac{1}{\lambda_{\cal R}}&=p\frac{\lambda^2}{\lambda_{\cal R}}
 \\
 &=2\sqrt{\frac{n}{(d-2)(d+n-2)}}=2\Big|\frac{\nabla_\phi \Lambda_{\rm sp}}{\kappa_d\Lambda_{\rm sp}}\Big|,\label{eq:Cond2}}
 where for the first line we use \eqref{eq:solB}, recalling that $p=\frac{4}{(d-2)\lambda^2}$ is   defined by the time dependence of the scale factor $a(t)\sim t^p$, while for the last line  \eqref{eq:KKLN} is used.
 \footnote{We also note that since $\lambda_{\cal R}\geq \frac{2}{\sqrt{d-2}}$, $p$ is bounded by $p=\frac{4}{(d-2)\lambda^2}\leq \frac{\lambda_{\cal R}^2}{\lambda^2}$.}

Suppose $\lambda \geq \lambda_{\cal R}$, that is, the dominant potential decays as or more rapidly than $V_{\cal R}$.
Since $\lambda_{\cal R}\geq 2/\sqrt{d-2}$, we infer from  \eqref{eq:RATIO} that $R_{\rm sp}/R_{\rm hor}\leq \lambda_{\cal R}/\lambda \leq 1$.
Therefore, in this case,  $R_{\rm sp}$ is always equal to or smaller than $R_{\rm hor}$ so $S_{\rm sp}$ cannot exceed  $S_{\rm hor}$  provided $S_{\rm sp}<S_{\rm hor}$ initially.
Meanwhile, if $\lambda<\lambda_{\cal R}$, the first line in \eqref{eq:Cond2} tells us that the inequality \eqref{eq:Cond1} can be written as $p>\lambda_{\cal R}/\lambda$, hence $p>1$.
As we have seen in section \ref{subsec:cosmo}, this is nothing more than the condition that the background geometry has an event horizon of the finite size or equivalently,  the trans-Planckian censorship bound is violated.
We also note that since the condition $p>1$ is equivalent to $\lambda<\frac{2}{\sqrt{d-2}}(\leq \lambda_{\cal R})$, the trans-Planckian censorship bound cannot be violated when $\lambda \geq \lambda_{\cal R}$.
  This shows that  when $R_{\rm sp}>R_{\rm hor}$, that is., the background is unstable in the entropic point of view, $p>1$ is satisfied, which means that the background admits the event horizon of finite size, violating the trans-Planckian censorship bound.
However, the converse is not always true : the presence of the event horizon   does not guarantee the relation $R_{\rm sp}>R_{\rm hor}$.
\footnote{We are grateful to the referee for pointing out this.} 
This can be found from the fact that $\lambda_{\cal R}\geq 2/\sqrt{d-2}$ implies $4/[(d-2)\lambda_{\cal R}]<2/\sqrt{d-2}$, which allows  the value of $\lambda$ satisfying
\dis{\frac{4}{(d-2)\lambda_{\cal R}}<\lambda <\frac{2}{\sqrt{d-2}},\label{eq:loophole}}
where the first and the second inequalities are equivalent to $R_{\rm sp}<R_{\rm hor}$ and $p>1$, respectively.
In other words, the background admitting the event horizon but entropically stable background can exist.
For example, when $d=4$ and $n=1$,  we infer from \eqref{eq:lR} that $\lambda_{\cal R}=\sqrt6$, hence $R_{\rm sp}>R_{\rm hor}$ is satisfied provided $\lambda<4/[(d-2)\lambda_{\cal R}]=\sqrt{2/3}$.
On the other hand, the condition for the existence of the event horizon   reads $\lambda<2/\sqrt{d-2}=\sqrt2$.
Then, for $\lambda \in (\sqrt{2/3}, \sqrt2)$, the background, which is stable but admits the event horizon is allowed.
Indeed, two conditions $R_{\rm sp}>R_{\rm hor}$ and $p>1$ are equivalent only in the limit $n \to \infty$, that is, the string tower limit, where $\lambda_{\cal R} \to 2/\sqrt{d-2}$ indicates that the value of $\lambda$ satisfying \eqref{eq:loophole} does not exist.
Therefore, we conclude that
\begin{quote}
In the quintessence model, the trans-Planckian censorship bound stating the absence of the event horizon implies  the condition  $R_{\rm sp}<R_{\rm hor}$, which means that $S_{\rm sp}=N_{\rm sp}$ does not exceed $S_{\rm hor}$ forever if $S_{\rm sp}<S_{\rm hor}$ initially.  
\end{quote} 
Whereas the converse is not true for the KK tower, in the case of the string tower, $\lambda_{\cal R}={2}/{\sqrt{d-2}}$, then the condition $R_{\rm sp}<R_{\rm hor}$ is equivalent to $\lambda_{\cal R}<\lambda$  (which is evident from  \eqref{eq:RATIO}), hence the converse is also true : the trans-Planckian censorship bound is equivalent to the background stability condition $R_{\rm sp}<R_{\rm hor}$.

 We also point out that when the background geometry is unstable by violating  the trans-Planckian censorship bound, the lifetime defined by the  time that takes for $S_{\rm sp}$ to saturate  $S_{\rm hor}$ can be estimated \cite{Seo:2019wsh, Cai:2019dzj}.
 Since the unstable background  arises in the solution B, $H$ is given by \eqref{eq:Hphi}, then we infer from   \eqref{eq:Shor} that 
   \dis{S_{\rm hor}=\frac{4\pi^{\frac{d+1}{2}}}{ \Gamma\big(\frac{d-1}{2}\big)}\Big[\frac{\lambda}{2}\sqrt{\frac{d-2}{d-1}\Big(\frac{4}{\lambda^2}\frac{d-1}{d-2}-1\Big)}\Big]^{d-2}\frac{e^{\frac{\lambda}{2}(d-2)\kappa_d\phi}}{\kappa_d^2 H_0^{d-2}},}
  where we define the fiducial Hubble parameter $H_0$ by
 \dis{H_0^2=\frac{2 \kappa_d^2}{(d-1)(d-2)}V_0.}
Then, rapidly increasing  $S_{\rm sp}=N_{\rm sp}=e^{\frac{2}{\lambda_{\cal R}}\kappa_d \phi}$ saturates $S_{\rm hor}$ when 
\dis{\kappa_d\phi = \frac{\lambda_{\cal R}}{2}\frac{1}{1-\frac{d-2}{4}\lambda \lambda_{\cal R}}\Big[\log\Big(\frac{1}{\kappa_d^2 H_0^{d-2}}\Big)+\log\Big(\frac{4\pi^{\frac{d+1}{2}}}{ \Gamma\big(\frac{d-1}{2}\big)}\Big[\frac{\lambda}{2}\sqrt{\frac{d-2}{d-1}\Big(\frac{4}{\lambda^2}\frac{d-1}{d-2}-1\Big)}\Big]^{d-2}\Big)\Big],}
which is positive provided \eqref{eq:Cond1} is satisfied, that is, $R_{\rm sp}>R_{\rm hor}$. 
Here the second term may be neglected when $1/\kappa_d^2 H_0^{d-2}\sim (M_{{\rm Pl},d}/H_0)^{d-2}$ is sufficiently large, which is typical.
 Using \eqref{eq:tphi} (or \eqref{eq:solB}), this can be rewritten as the time scale,
 \dis{t\sim \frac{2}{\lambda H_0}\sqrt{\frac{\big(\frac{4}{\lambda^2}\frac{d-1}{d-2}-1\big)}{(d-1)(d-2)}} \Big(\frac{1}{\kappa_d^2 H_0^{d-2}}\Big)^{\frac{\lambda\lambda_{\cal R}}{4}\frac{1}{1-\frac{d-2}{4}\lambda\lambda_{\cal R}}}.\label{eq:lifttime}}
 In particular, in the limit $\lambda \to 0$, it can be approximated as
 \dis{t\sim \frac{1}{d-2}\frac{\lambda_{\cal R}}{\lambda}\frac{1}{H_0}\log\Big[\frac{1}{\kappa_d^2 H_0^{d-2}}\Big]+{\rm constant},}
 as claimed by the trans-Planckian censorship conjecture.
Here the prefactor $1/\lambda$ can be understood as follows.
For the solution B given by \eqref{eq:solB}, we obtain
\dis{\kappa_d \dot{\phi}=\frac{2}{\lambda}\frac{1}{t},\quad\quad 
H=\frac{4}{(d-2)\lambda^2}\frac{1}{t},}
then the slow-roll parameter is (see also \eqref{eq:EoM})
\dis{\epsilon_H=-\frac{\dot H}{H^2}=\frac{\kappa_d^2\dot{\phi}^2}{(d-2)H^2}=\frac{d-2}{4}\lambda^2,}
which suggests that \eqref{eq:lifttime} can be rewritten as $t\sim (\sqrt{\epsilon_H} H_0)^{-1}\log(M_{{\rm Pl},d}/H_0)$.
Here $\sqrt{\epsilon_H}$ naturally appears when we convert $\phi$ to $\langle \dot{\phi}\rangle_t t$ since it can be interpreted as a factor proportional to   $\dot{\phi}$, as observed in \cite{Seo:2019wsh, Cai:2019dzj}. 

   \subsection{Implication on the scale separation}
   \label{Sec:scale} 
 
One of issues in the swampland program concerning cosmology is the scale separation between $m_{\rm KK}$ and $H$, which is motivated by the difficulty in obtaining the reliable   AdS background solution with  $m_{\rm KK} \gg H$ in the string theory model \cite{Gibbons:1985, deWit:1986mwo, Maldacena:2000mw, Gautason:2015tig} (for a review, see \cite{Coudarchet:2023mfs}).
It may be interpreted as the stability of the $d$-dimensional EFT against the appearance of the extra dimensions.
 That is, since $H$ is regarded as the IR cutoff, if $m_{\rm KK}$ decreases more rapidly than $H$, the EFT in the  limit $\phi \to \infty$ must find extra dimensions, hence  is no longer the $d$-dimensional one.
\footnote{Since the string tower corresponds to the $n\to \infty$ limit of the KK tower, we may discuss the scale separation between $M_s$ and $H$ in the similar manner, which is interpreted as the stability of the point particle description.} 
Regarding the scale separation in the quintessence model, it was pointed out in \cite{Andriot:2025cyi, Bedroya:2025ris} that $m_{\rm KK}$ can be well above $H$  when the background   evolves in time, rather than being stabilized.
Since two entropies, $S_{\rm sp}$ and $S_{\rm hor}$, we have compared are associated with two scales, $m_t$ and $H$, respectively, one can find the condition that the quintessence model admits the scale separation by comparing the scaling behavior of entropies with respect to the scalar field value.

  More concretely, for the KK tower, we infer from \eqref{eq:mkkVr} that 
  \dis{\frac{\nabla_\phi m_{\rm KK}}{\kappa_d m_{\rm KK}}=-\sqrt{\frac{d+n-2}{n(d-2)}}=-\frac{\lambda_{\cal R}}{2}.}
This will be compared with the decreasing rate of $H$.  
  For the solution A where $H\sim {\rm exp}\big[-\sqrt{\frac{d-1}{d-2}}\kappa_d \phi\big]$ is satisfied (see \eqref{eq:HphiA}), the decreasing rate of $H$   is given by $\nabla_\phi H/(\kappa_d H)=-\sqrt{\frac{d-1}{d-2}}$.
  Since $\lambda_{\cal R} \leq 2\sqrt{\frac{d-1}{d-2}}$, we conclude that $H$   decreases more rapidly compared to $m_{\rm KK}$, hence the scale separation can be achieved.
 On the other hand, for the solution B,  we obtain from  $H \sim {\rm exp}\big[-\frac{\lambda}{2}\kappa_d \phi\big]$ (see \eqref{eq:Hphi}) that
\dis{\frac{\nabla_\phi H}{\kappa_d H}=-\frac{\lambda}{2}.}
Therefore, if $\lambda <\lambda_{\cal R}$, $m_{\rm KK}$ decreases more rapidly than $H$, then in the limit $\phi\to \infty$  (or equivalently, $t\to\infty$), $m_{\rm KK}$ cannot be much above $H$.
In other words, the scale separation is not achieved.
This includes the case where the trans-Planckian censorship bound is violated : the value of $\lambda$ in this case is  smaller than $\frac{2}{\sqrt{d-2}}$, which corresponds to the lower bound on $\lambda_{\cal R}$.
 
 In terms of the decreasing rates of the entropies, the scale separation condition for the solution B reads  
 \dis{\frac{R_{\rm sp} R_{\rm hor}}{d-2}=\frac{\lambda}{\lambda_{\cal R}}\geq 1,\label{Eq:Rprod}}
where a factor $d-2$ reflects the fact that $S_{\rm sp}$ obeys the area law.
  This can be rewritten as
\dis{\frac{\nabla  \sqrt{V}}{\kappa_d \sqrt{V}}\cdot\frac{\nabla\Lambda_{\rm sp}}{\kappa_d \Lambda_{\rm sp}} =\frac{\lambda}{2}\times \Big(\frac{2}{d-2}\frac{1}{\lambda_{\cal R}}\Big) \geq \frac{1}{d-2}=\frac{\nabla  m_{\rm KK}}{\kappa_d m_{\rm KK}}\cdot\frac{\nabla\Lambda_{\rm sp}}{\kappa_d \Lambda_{\rm sp}},}
which is the form considered in \cite{Bedroya:2025ltj}, explicitly showing that the decreasing rate of $H\propto  \sqrt{V}$ is larger than that of $m_{\rm KK}$.
Meanwhile, \eqref{Eq:Rprod} can be interpreted similarly to   the uncertainty principle : for the model achieving the scale separation, the product $R_{\rm sp}R_{\rm hor}$ cannot be smaller than the constant $d-2$ hence the smaller $R_{\rm sp}$ is the larger  $R_{\rm hor}$   and vice versa.
\footnote{It was pointed out in \cite{Anchordoqui:2025izb} that under the quantization $[\phi, \dot{\phi}]=i$, which reflects the uncertainty principle, two functions $f(\phi)$ and $h(\phi)$ of $\phi$  satisfy $[f(\phi), \dot{h}(\phi)]=i\nabla_\phi f \nabla_\phi h$, hence $\Delta f(\phi) \Delta \dot{h}(\phi)  \geq \frac12|\nabla_\phi f \nabla_\phi h|$. 
This indicates that the product of decreasing(or increasing) rates we discussed can be interpreted as the bound imposed by the uncertainty principle.
This must be distinguished from our analogue uncertainty principle, which considers the lower bound on $\nabla_\phi f \nabla_\phi h$ itself. }
Of course, the additional condition of the stability of the background    excludes the case $R_{\rm sp} > R_{\rm hor}$.
We also note  that the bound \eqref{Eq:Rprod} is saturated when $\lambda=\lambda_{\cal R}$, that is,   $V$ has the same scaling behavior as $V_{\cal R}$ with respect to $\phi$, which includes  $V=V_{\cal R}$.
 As discussed in section \ref{Sec:instab}, when $\lambda=\lambda_{\cal R}$, the ratio $R_{\rm sp}/R_{\rm hor}=4/[(d-2)\lambda_{\cal R}^2]$ is always smaller than $1$ for the KK tower, hence the background   is stable.

 The scale separation bound \eqref{Eq:Rprod} also can be compared with the AdS distance conjecture, another conjecture concerning the scale separation \cite{Lust:2019zwm}.
 It claims that the tower mass scale and $V$ (the vacuum energy) are connected by the scaling behavior $m_t \sim V^\alpha$, and the exponent $\alpha$ cannot be larger than $1/2$.
 This bound is motivated by the  relation $m_{t}\sim V^{1/2}$ for the supersymmetric AdS vacuum and also the Higuchi bound \cite{Higuchi:1986py} of the dS background. 
   For the quintessence model, the exponent is  given by $\alpha=\lambda_{\cal R}/(2\lambda)$.
 Therefore, the AdS distance conjecture bound reads $\lambda_{\cal R}/\lambda \leq 1$, which coincides with the condition for the scale separation given by \eqref{Eq:Rprod}.
 Moreover, it was suggested that the lower bound on $\alpha$ is $1/d$ \cite{Rudelius:2021oaz, Castellano:2021mmx, Montero:2022prj}, which reflects the fact that the 1-loop correction of the potential is at least $m_t^d$, hence $\lambda \leq \frac{d}{2}\lambda_{\cal R}$.
 \footnote{The same bound also can be obtained from the Cohen-Kaplan-Nelson (CKN) bound \cite{Cohen:1998zx}, which forbids the black hole formation at energy scale below the UV cutoff \cite{Castellano:2021mmx}.  }
 While discussion in section \ref{Sec:instab} shows that the background   is stable  against the descent of a tower of states provided $\lambda>\frac{4}{d-2}\frac{1}{\lambda_{\cal R}}$, since $\frac{4}{d-2}\frac{1}{\lambda_{\cal R}} \leq \frac{d}{2}\lambda_{\cal R}$ is always satisfied, the lower bound on $\alpha$ in the AdS distance conjecture is consistent with the stability of the background   if
 \dis{\frac{4}{d-2}\frac{1}{\lambda_{\cal R}} <\lambda \leq \frac{d}{2}\lambda_{\cal R}.}

 \section{Conclusions}
\label{sec:conclusion}

In this article, we investigate the instability condition of the background  in the quintessence model by applying the covariant entropy bound argument for the dS swampland conjecture.
That is,  we impose that the background  is stable if the increasing rate of the geometrical entropy proportional to the apparent horizon area is larger than that of the species entropy in the large scalar field limit.
  Then, one finds that the instability condition implies the condition that  the trans-Planckian censorship bound is violated, and the converse is also true for the string tower. 
The unstable background in this case corresponds to the accelerating expansion of the universe with a finite size of the event horizon, which is realized by  the nearly flat potential with decreasing rate $\lambda<\frac{4}{d-2}\frac{1}{\lambda_{\cal R}}$, where $\lambda_{\cal R}$ is the decreasing rate of the potential generated by the Ricci scalar of the extra dimensions.
Indeed, the  lifetime of the unstable background is also consistent with the trans-Planck censorship conjecture.
On the other hand, the scale separation condition, that is, the condition that the KK mass scale is kept higher than the Hubble parameter even in the large scalar field limit, can be written as the lower bound given by $d-2$ on the product of the decreasing rates of the geometrical and the species entropies.
This bound is consistent with the scaling behavior of the tower mass scale with respect to the vacuum energy proposed by the AdS distance conjecture.

Our discussion suggests the possibility that various swampland conjectures can be comprehensively understood in light of the entropy.
This not just reveals the connection between conjectures, but also enables us to compare the swampland program with thermodynamic  description of quantum gravity, which has been developed extensively by studying, for example, black hole.

\subsection*{Data availability Statement}

Data sharing not applicable to this article as no datasets were generated or analyzed during the current study.


%


\appendix


\renewcommand{\theequation}{\Alph{section}.\arabic{equation}}



\begin{thebibliography}{99}

\small

\bibitem{Vafa:2005ui}
C.~Vafa,
[arXiv:hep-th/0509212 [hep-th]].

\bibitem{Brennan:2017rbf}
T.~D.~Brennan, F.~Carta and C.~Vafa,
PoS \textbf{TASI2017}, 015 (2017)
[arXiv:1711.00864 [hep-th]].

\bibitem{Palti:2019pca}
E.~Palti,
Fortsch. Phys. \textbf{67}, no.6, 1900037 (2019)
[arXiv:1903.06239 [hep-th]].

\bibitem{vanBeest:2021lhn}
M.~van Beest, J.~Calder{\'o}n-Infante, D.~Mirfendereski and I.~Valenzuela,
Phys. Rept. \textbf{989}, 1-50 (2022)
[arXiv:2102.01111 [hep-th]].

\bibitem{Grana:2021zvf}
M.~Gra{\~n}a and A.~Herr{\'a}ez,
Universe \textbf{7}, no.8, 273 (2021)
[arXiv:2107.00087 [hep-th]].

\bibitem{Agmon:2022thq}
N.~B.~Agmon, A.~Bedroya, M.~J.~Kang and C.~Vafa,
[arXiv:2212.06187 [hep-th]].

\bibitem{Danielsson:2018ztv}
U.~H.~Danielsson and T.~Van Riet,
Int. J. Mod. Phys. D \textbf{27}, no.12, 1830007 (2018)

[arXiv:1804.01120 [hep-th]].

\bibitem{Obied:2018sgi}
G.~Obied, H.~Ooguri, L.~Spodyneiko and C.~Vafa,
[arXiv:1806.08362 [hep-th]].

\bibitem{Andriot:2018wzk}
D.~Andriot,
Phys. Lett. B \textbf{785}, 570-573 (2018)
[arXiv:1806.10999 [hep-th]].

\bibitem{Garg:2018reu}
S.~K.~Garg and C.~Krishnan,
JHEP \textbf{11}, 075 (2019)
[arXiv:1807.05193 [hep-th]].

\bibitem{Cicoli:2018kdo}
M.~Cicoli, S.~De Alwis, A.~Maharana, F.~Muia and F.~Quevedo,
Fortsch. Phys. \textbf{67}, no.1-2, 1800079 (2019)
[arXiv:1808.08967 [hep-th]].


\bibitem{Ooguri:2018wrx}
H.~Ooguri, E.~Palti, G.~Shiu and C.~Vafa,
Phys. Lett. B \textbf{788}, 180-184 (2019)
[arXiv:1810.05506 [hep-th]].

\bibitem{Andriot:2018mav}
D.~Andriot and C.~Roupec,
Fortsch. Phys. \textbf{67}, no.1-2, 1800105 (2019)
[arXiv:1811.08889 [hep-th]].

\bibitem{Bousso:1999xy}
R.~Bousso,
JHEP \textbf{07}, 004 (1999)
[arXiv:hep-th/9905177 [hep-th]].

\bibitem{Bousso:2002ju}
R.~Bousso,
Rev. Mod. Phys. \textbf{74}, 825-874 (2002)
[arXiv:hep-th/0203101 [hep-th]].

\bibitem{Ooguri:2006in}
H.~Ooguri and C.~Vafa,
Nucl. Phys. B \textbf{766}, 21-33 (2007)
[arXiv:hep-th/0605264 [hep-th]].


\bibitem{Lee:2019xtm}
S.~J.~Lee, W.~Lerche and T.~Weigand,
JHEP \textbf{02}, 096 (2022)
[arXiv:1904.06344 [hep-th]].

\bibitem{Lee:2019wij}
S.~J.~Lee, W.~Lerche and T.~Weigand,
JHEP \textbf{02}, 190 (2022)
[arXiv:1910.01135 [hep-th]].

\bibitem{Padmanabhan:2012ik}
T.~Padmanabhan,
[arXiv:1206.4916 [hep-th]].

\bibitem{Hadi:2019qxn}
H.~Hadi, F.~Darabi and Y.~Heydarzade,
EPL \textbf{131}, no.5, 59001 (2020)
[arXiv:1907.07143 [gr-qc]].



\bibitem{Bedroya:2019snp}
A.~Bedroya and C.~Vafa,
JHEP \textbf{09}, 123 (2020)
[arXiv:1909.11063 [hep-th]].


\bibitem{Peebles:1987ek}
P.~J.~E.~Peebles and B.~Ratra,
Astrophys. J. Lett. \textbf{325}, L17 (1988)

\bibitem{Ratra:1987rm}
B.~Ratra and P.~J.~E.~Peebles,
Phys. Rev. D \textbf{37}, 3406 (1988)

\bibitem{Caldwell:1997ii}
R.~R.~Caldwell, R.~Dave and P.~J.~Steinhardt,
Phys. Rev. Lett. \textbf{80}, 1582-1585 (1998)
[arXiv:astro-ph/9708069 [astro-ph]].

\bibitem{Martin:2008qp}
J.~Martin,
Mod. Phys. Lett. A \textbf{23}, 1252-1265 (2008)
[arXiv:0803.4076 [astro-ph]].

\bibitem{Tsujikawa:2013fta}
S.~Tsujikawa,
Class. Quant. Grav. \textbf{30}, 214003 (2013)
[arXiv:1304.1961 [gr-qc]].

\bibitem{Hellerman:2001yi}
S.~Hellerman, N.~Kaloper and L.~Susskind,
JHEP \textbf{06}, 003 (2001)
[arXiv:hep-th/0104180 [hep-th]].

\bibitem{Fischler:2001yj}
W.~Fischler, A.~Kashani-Poor, R.~McNees and S.~Paban,
JHEP \textbf{07}, 003 (2001)
[arXiv:hep-th/0104181 [hep-th]].

\bibitem{Kaloper:2008qs}
N.~Kaloper and L.~Sorbo,
Phys. Rev. D \textbf{79}, 043528 (2009)
[arXiv:0810.5346 [hep-th]].

\bibitem{Cicoli:2012tz}
M.~Cicoli, F.~G.~Pedro and G.~Tasinato,
JCAP \textbf{07}, 044 (2012)
[arXiv:1203.6655 [hep-th]].

\bibitem{Olguin-Trejo:2018zun}
Y.~Olguin-Trejo, S.~L.~Parameswaran, G.~Tasinato and I.~Zavala,
JCAP \textbf{01}, 031 (2019)
[arXiv:1810.08634 [hep-th]].

\bibitem{Cicoli:2021fsd}
M.~Cicoli, F.~Cunillera, A.~Padilla and F.~G.~Pedro,
Fortsch. Phys. \textbf{70}, no.4, 2200009 (2022)
[arXiv:2112.10779 [hep-th]].

\bibitem{Cicoli:2021skd}
M.~Cicoli, F.~Cunillera, A.~Padilla and F.~G.~Pedro,
Fortsch. Phys. \textbf{70}, no.4, 2200008 (2022)
[arXiv:2112.10783 [hep-th]].

\bibitem{Brinkmann:2022oxy}
M.~Brinkmann, M.~Cicoli, G.~Dibitetto and F.~G.~Pedro,
JHEP \textbf{11}, 044 (2022)
[arXiv:2206.10649 [hep-th]].
 
\bibitem{Conlon:2022pnx}
J.~P.~Conlon and F.~Revello,
JHEP \textbf{11}, 155 (2022)
[arXiv:2207.00567 [hep-th]].

\bibitem{Rudelius:2022gbz}
T.~Rudelius,
JHEP \textbf{10}, 018 (2022)
[arXiv:2208.08989 [hep-th]].

\bibitem{Calderon-Infante:2022nxb}
J.~Calder{\'o}n-Infante, I.~Ruiz and I.~Valenzuela,
JHEP \textbf{06}, 129 (2023)
[arXiv:2209.11821 [hep-th]].
 
\bibitem{Apers:2022cyl}
F.~Apers, J.~P.~Conlon, M.~Mosny and F.~Revello,
JHEP \textbf{08}, 156 (2023)
[arXiv:2212.10293 [hep-th]].
 
\bibitem{Shiu:2023nph}
G.~Shiu, F.~Tonioni and H.~V.~Tran,
Phys. Rev. D \textbf{108}, no.6, 063527 (2023)
[arXiv:2303.03418 [hep-th]].

\bibitem{Shiu:2023fhb}
G.~Shiu, F.~Tonioni and H.~V.~Tran,
Phys. Rev. D \textbf{108}, no.6, 063528 (2023)
[arXiv:2306.07327 [hep-th]].

\bibitem{Seo:2024fki}
M.~S.~Seo,
Nucl. Phys. B \textbf{1008}, 116705 (2024)
[arXiv:2402.00241 [hep-th]].

\bibitem{Seo:2024qzf}
M.~S.~Seo,
Fortsch. Phys. \textbf{72}, no.11, 2400112 (2024)
[arXiv:2403.07307 [hep-th]].


\bibitem{Bedroya:2022tbh}
A.~Bedroya,
JHEP \textbf{06}, 016 (2024)
[arXiv:2211.09128 [hep-th]].

\bibitem{Bedroya:2025fie}
A.~Bedroya and P.~J.~Steinhardt,
[arXiv:2511.15784 [hep-th]].


\bibitem{Strominger:2001pn}
A.~Strominger,
JHEP \textbf{10}, 034 (2001)
[arXiv:hep-th/0106113 [hep-th]].
 
\bibitem{Seo:2019wsh}
M.~S.~Seo,
Phys. Lett. B \textbf{807}, 135580 (2020)
[arXiv:1911.06441 [hep-th]].

\bibitem{Cai:2019dzj}
R.~G.~Cai and S.~J.~Wang,
Sci. China Phys. Mech. Astron. \textbf{64}, no.1, 210011 (2021)
[arXiv:1912.00607 [hep-th]].



\bibitem{Sun:2019obt}
S.~Sun and Y.~L.~Zhang,
Phys. Lett. B \textbf{816}, 136245 (2021)
[arXiv:1912.13509 [hep-th]].

\bibitem{Cribiori:2025oek}
N.~Cribiori and F.~Tonioni,
JHEP \textbf{02}, 035 (2026)
[arXiv:2507.02738 [hep-th]].

\bibitem{Cribiori:2023ffn}
N.~Cribiori, D.~Lust and C.~Montella,
JHEP \textbf{10}, 059 (2023)
[arXiv:2305.10489 [hep-th]].

\bibitem{Basile:2023blg}
I.~Basile, D.~L{\"u}st and C.~Montella,
JHEP \textbf{07}, 208 (2024)
[arXiv:2311.12113 [hep-th]].

\bibitem{Basile:2024dqq}
I.~Basile, N.~Cribiori, D.~Lust and C.~Montella,
JHEP \textbf{06}, 127 (2024)
[arXiv:2401.06851 [hep-th]].

\bibitem{Herraez:2024kux}
A.~Herr{\'a}ez, D.~L{\"u}st, J.~Masias and M.~Scalisi,
SciPost Phys. \textbf{18}, 083 (2025)
[arXiv:2406.17851 [hep-th]].

\bibitem{Herraez:2025clp}
A.~Herr{\'a}ez, D.~L{\"u}st, J.~Masias and C.~Montella,
PoS \textbf{CORFU2024}, 161 (2025)
[arXiv:2506.02335 [hep-th]].

\bibitem{Coudarchet:2023mfs}
T.~Coudarchet,
Phys. Rept. \textbf{1064}, 1-28 (2024)
[arXiv:2311.12105 [hep-th]].

\bibitem{Veneziano:2001ah}
G.~Veneziano,
JHEP \textbf{06}, 051 (2002)
[arXiv:hep-th/0110129 [hep-th]].

\bibitem{Dvali:2007hz}
G.~Dvali,
Fortsch. Phys. \textbf{58}, 528-536 (2010)
[arXiv:0706.2050 [hep-th]].

\bibitem{Dvali:2007wp}
G.~Dvali and M.~Redi,
Phys. Rev. D \textbf{77}, 045027 (2008)
[arXiv:0710.4344 [hep-th]].

\bibitem{Dvali:2009ks}
G.~Dvali and D.~Lust,
Fortsch. Phys. \textbf{58}, 505-527 (2010)
[arXiv:0912.3167 [hep-th]].

\bibitem{Dvali:2010vm}
G.~Dvali and C.~Gomez,
[arXiv:1004.3744 [hep-th]].

\bibitem{Dvali:2012uq}
G.~Dvali, C.~Gomez and D.~Lust,
Fortsch. Phys. \textbf{61}, 768-778 (2013)
[arXiv:1206.2365 [hep-th]].

\bibitem{Kani:1989im}
I.~Kani and C.~Vafa,
Commun. Math. Phys. \textbf{130}, 529-580 (1990)

\bibitem{Castellano:2021mmx}
A.~Castellano, A.~Herr{\'a}ez and L.~E.~Ib{\'a}{\~n}ez,
JHEP \textbf{08}, 217 (2022)
[arXiv:2112.10796 [hep-th]].

\bibitem{Castellano:2023stg}
A.~Castellano, I.~Ruiz and I.~Valenzuela,
Phys. Rev. Lett. \textbf{132}, no.18, 181601 (2024)
[arXiv:2311.01501 [hep-th]].

\bibitem{Castellano:2023jjt}
A.~Castellano, I.~Ruiz and I.~Valenzuela,
JHEP \textbf{06}, 037 (2024)
[arXiv:2311.01536 [hep-th]].

\bibitem{Lucchin:1984yf}
F.~Lucchin and S.~Matarrese,
Phys. Rev. D \textbf{32}, 1316 (1985)

\bibitem{Copeland:1997et}
E.~J.~Copeland, A.~R.~Liddle and D.~Wands,
Phys. Rev. D \textbf{57}, 4686-4690 (1998)
[arXiv:gr-qc/9711068 [gr-qc]].
 
\bibitem{vandenHoogen:1999qq}
R.~J.~van den Hoogen, A.~A.~Coley and D.~Wands,
Class. Quant. Grav. \textbf{16}, 1843-1851 (1999)
[arXiv:gr-qc/9901014 [gr-qc]].

\bibitem{Seo:2019mfk}
M.~S.~Seo,
Phys. Lett. B \textbf{797}, 134904 (2019)
[arXiv:1907.12142 [hep-th]].
 
\bibitem{Seo:2022uaz}
M.~S.~Seo,
JCAP \textbf{11}, 005 (2022)
[arXiv:2206.05857 [hep-th]].

\bibitem{vandeHeisteeg:2022btw}
D.~van de Heisteeg, C.~Vafa, M.~Wiesner and D.~H.~Wu,
Beijing J. Pure Appl. Math. \textbf{1}, no.1, 1-41 (2024)
[arXiv:2212.06841 [hep-th]].


\bibitem{Cribiori:2022nke}
N.~Cribiori, D.~L{\"u}st and G.~Staudt,
Phys. Lett. B \textbf{844}, 138113 (2023)
[arXiv:2212.10286 [hep-th]].


\bibitem{Hayward:1997jp}
S.~A.~Hayward,
Class. Quant. Grav. \textbf{15}, 3147-3162 (1998)
[arXiv:gr-qc/9710089 [gr-qc]].

\bibitem{Bak:1999hd}
D.~Bak and S.~J.~Rey,
Class. Quant. Grav. \textbf{17}, L83 (2000)
[arXiv:hep-th/9902173 [hep-th]].

\bibitem{Cai:2005ra}
R.~G.~Cai and S.~P.~Kim,
JHEP \textbf{02}, 050 (2005)
[arXiv:hep-th/0501055 [hep-th]].

\bibitem{Cai:2006rs}
R.~G.~Cai and L.~M.~Cao,
Phys. Rev. D \textbf{75}, 064008 (2007)
[arXiv:gr-qc/0611071 [gr-qc]].

\bibitem{Abreu:2010ru}
G.~Abreu and M.~Visser,
Phys. Rev. D \textbf{82}, 044027 (2010)
[arXiv:1004.1456 [gr-qc]].

\bibitem{Binetruy:2014ela}
P.~Bin{\'e}truy and A.~Helou,
Class. Quant. Grav. \textbf{32}, no.20, 205006 (2015)
[arXiv:1406.1658 [gr-qc]].

\bibitem{Cohen:1998zx}
A.~G.~Cohen, D.~B.~Kaplan and A.~E.~Nelson,
Phys. Rev. Lett. \textbf{82}, 4971-4974 (1999)
[arXiv:hep-th/9803132 [hep-th]].

\bibitem{Banks:2019arz}
T.~Banks and P.~Draper,
Phys. Rev. D \textbf{101}, no.12, 126010 (2020)
[arXiv:1911.05778 [hep-th]].

\bibitem{Seo:2021bpb}
M.~S.~Seo,
Eur. Phys. J. C \textbf{82}, no.4, 338 (2022)
[arXiv:2106.00138 [hep-th]].



\bibitem{Gibbons:1985}
G.~Gibbons, 
''Aspects of Supergravity Theories", in Supersymmetry, Supergravity and Related Topics: Proceedings of the XVth GIFT International Seminar on Theoretical Physics, 4–9 June
1984, Sant Feliu de Gu\'ixols, Girona, Spain F. del Aguila, J. A. de Azc\'arraga and L. E. Ib\'a\~nez eds., World Scientific (1985) [ISBN: 9789971966805]

\bibitem{deWit:1986mwo}
B.~de Wit, D.~J.~Smit and N.~D.~Hari Dass,
Nucl. Phys. B \textbf{283}, 165 (1987)
doi:10.1016/0550-3213(87)90267-7



\bibitem{Maldacena:2000mw}
J.~M.~Maldacena and C.~Nunez,
Int. J. Mod. Phys. A \textbf{16}, 822-855 (2001)
[arXiv:hep-th/0007018 [hep-th]].

\bibitem{Gautason:2015tig}
F.~F.~Gautason, M.~Schillo, T.~Van Riet and M.~Williams,
JHEP \textbf{03}, 061 (2016)
[arXiv:1512.00457 [hep-th]].

\bibitem{Andriot:2025cyi}
D.~Andriot, N.~Cribiori and T.~Van Riet,
Phys. Rev. D \textbf{112}, no.2, 026028 (2025)
[arXiv:2504.08634 [hep-th]].

\bibitem{Bedroya:2025ris}
A.~Bedroya, H.~Lee and P.~Steinhardt,
[arXiv:2504.13260 [hep-th]].



\bibitem{Bedroya:2025ltj}
A.~Bedroya and P.~J.~Steinhardt,
[arXiv:2509.25313 [hep-th]].


\bibitem{Anchordoqui:2025izb}
L.~A.~Anchordoqui, D.~Lust and S.~L{\"u}st,
[arXiv:2510.25846 [hep-th]].


\bibitem{Lust:2019zwm}
D.~L{\"u}st, E.~Palti and C.~Vafa,
Phys. Lett. B \textbf{797}, 134867 (2019)
[arXiv:1906.05225 [hep-th]].

\bibitem{Higuchi:1986py}
A.~Higuchi,
Nucl. Phys. B \textbf{282}, 397-436 (1987)

\bibitem{Rudelius:2021oaz}
T.~Rudelius,
JHEP \textbf{08}, 041 (2021)
[arXiv:2101.11617 [hep-th]].


\bibitem{Montero:2022prj}
M.~Montero, C.~Vafa and I.~Valenzuela,
JHEP \textbf{02}, 022 (2023)
[arXiv:2205.12293 [hep-th]].




\end{thebibliography}
\end{document}